\documentclass[]{aastex63}
\usepackage{amsmath}

\graphicspath{{./}{figures/}}

\received{--}
\revised{--}
\accepted{--}
\submitjournal{ApJ}

\shorttitle{Meridional Circulation}
\shortauthors{Stejko et al.}

\begin{document}

\title{Forward Modeling Helioseismic Signatures of One- and Two-Cell Meridional Circulation}


\correspondingauthor{Andrey Stejko}
\email{ams226@njit.edu}

\author{Andrey M. Stejko}
\affiliation{New Jersey Institute of Technology\\
 323 Dr Martin Luther King Jr Blvd.\\
 Newark, NJ 07012, USA}

\author{Alexander G. Kosovichev}
\affiliation{New Jersey Institute of Technology\\
 323 Dr Martin Luther King Jr Blvd.\\
 Newark, NJ 07012, USA}

\author{Valery V. Pipin}
\affiliation{Institute of Solar-Terrestrial Physics\\  Russian Academy of Sciences\\
 Irkutsk, 664033, Russia}
 
\begin{abstract}

\indent Using a 3D global solver of the linearized Euler equations, we model acoustic oscillations over background velocity flow fields of single-cell meridional circulation with deep and shallow return flows as well as double-cell meridional circulation with strong and weak reversals. The velocities are generated using a mean-field hydrodynamic and dynamo model---moving through the regimes with minimal parameter changes; counter-rotation near the base of the tachocline is induced by sign inversion of the non-diffusive action of turbulent Reynolds stresses ($\Lambda$-effect) due to the radial inhomogeneity of the Coriolis number. By mimicking the stochastic excitation of resonant modes in the convective interior, we simulate realization noise present in solar observations. Using deep-focusing to analyze differences in travel-time signatures between the four regimes, as well as comparing to solar observations, we show that current helioseismology techniques may offer important insights about the location and strength of the return flow, however, that it may not currently be possible to definitively distinguish between profiles of single-cell or double-cell meridional circulation.

\end{abstract}



\section{Introduction}\label{sec:intro}

\indent Meridional mass flows are characterized by large circulation structures, redistributing angular momentum and magnetic flux \citep{2003ApJ...589..665H} throughout the convective interior of the Sun. Doppler measurements have been used to study the surface expressions of these circulation cells in detail \citep{1979SoPh...63....3D,1996ApJ...460.1027H,2010ApJ...725..658U}, showing strong poleward flows ($20\text{ m s}^{-1}$) in each hemisphere. The Michelson Doppler Imager (MDI) \citep{1995SoPh..162..129S} of the Solar and Heliospheric Observatory (SOHO) \citep{1995SoPh..162....1D}, its successor---the Helioseismic Magnetic Imager (HMI) \citep{2012SoPh..275..207S} aboard the Solar Data Observatory (SDO) spacecraft \citep{2012SoPh..275....3P}, as well as the Global Oscillation Network Group (GONG) \citep{1996Sci...272.1284H} have been instrumental in providing the long term observational data needed to analyze these structures.\\
\indent Applying techniques in local helioseismology to these observations allows us to probe below the solar surface \citep{2002RvMP...74.1073C,2005LRSP....2....6G}; cross-correlating measurements of surface oscillations provides insight into the structure of flows that advect acoustic rays resonating throughout the convective interior. While measurements in subsurface layers ($>0.96 R_{\odot}$) have remained relatively consistent \citep[e.g.][]{1997Natur.390...52G,2004ApJ...603..776Z,2015SoPh..290.3113K,2015ApJ...805..133J,2015ApJ...807..125B,2017ApJ...845....2B,2018ApJ...860...48L}, complications arise when attempting to probe deeper into the solar interior. The return flow of the circulation cell has historically been theorized to sit near the tachocline ($\sim 0.72\text{ R}_{\odot}$), with a velocity weaker than that of the surface flow by an order of magnitude ($\sim 2\text{ m s}^{-1}$) \citep{1999_giles}---the minimal impact of these flows lead to difficulties in clearly resolving deep meridional circulation structures. These problems are compounded by systematic errors such as the center-to-limb (CtoL) variations in travel-time measurements \citep{2012ApJ...749L...5Z}---the physical basis of which has yet to be fully understood. This effect is an order of magnitude larger than that of meridional circulation signatures, magnifying the impact of any errors made when attempting to estimate it. New empirical approaches in disentangling this effect using frequency-dependent analysis, however, lend confidence to our ability to remove this error from solar measurements \citep[e.g.][]{2018ApJ...853..161C,2020arXiv200412708R}. Another systematic error in helioseismology measurements can result from the interference of what appear to be effective downflows in surface magnetic regions (see \citet{2015ApJ...805..165L}). In order to avoid complications from these regions they can be masked out \citep{2015ApJ...805..165L,2017ApJ...849..144C,2018ApJ...853..161C,2018ApJ...860...48L,2019_chen,2020Sci...368.1469G}, however this results in a reduction of available observational data. Correlation signals are strongly obscured by a realization noise \citep{2004ApJ...614..472G} resulting from turbulent convection near the solar surface. The confidence of any inference of internal solar structures is based on the bounds that we can place on this noise---a systematic examination of which we explore in this paper. \\ 
\indent The manifestation of these problems can be seen in the variability of inferred meridional velocity profiles, with more recent estimates of much shallower return flow depths (from $\sim 40$ to $\sim 70$ \text{ Mm}; \cite{2007AN....328.1009M, 2012ApJ...760...84H} respectively). The nature of the single-cell model has also been put into question with a helioseismic analysis of MDI, HMI and GONG data \citep{2013ApJ...774L..29Z,2013ApJ...778L..38S,2014ApJ...784..145K,2018IAUS..340...13B,2018ApJ...860...48L,2019_chen} inferring a two-cell or even a multi-cell structure---a result whose viability has been validated by numerical, convectively driven, MHD/HD simulations \citep{2002ApJ...570..865B,2006ApJ...641..618M,2013ApJ...779..176G,2019ApJ...871..217M}. Recent analysis of MDI and GONG data, however, continues to insist on the single-cell model \citep{2020Sci...368.1469G}---typically employed by mean-field simulations of the Sun \citep{2005ApJ...622.1320R,2011ApJ...740...12H,2011A&A...530A..48K,2013IAUS..294..399K} and other late-type stars \citep{2012MNRAS.423.3344K}.\\
\indent The transition between single- and double-cell circulation structures has been explored in conjunction with anti-Solar and Solar profiles of differential rotation in the convectively driven hydrodynamic models of \citet{2013ApJ...779..176G,2014MNRAS.438L..76G,2014A&A...570A..43K,2015ApJ...804...67F}. These simulations highlight the impact of angular momentum transport by non-diffusive turbulent Reynolds stresses via gyroscopic pumping (the $\Lambda$-effect, \citet{1989drsc.book.....R, 1993A&A...276...96K}). Mean-field modeling has shown that altering this parameter ($\Lambda$) \citep{2017ApJ...835....9B} can induce the formation of counter-rotation cells in the lower convection zone, with \citet{2018ApJ...854...67P} demonstrating that radial variations of the Coriolis number near the tachocline can be a potential physical basis for inducing a sign inversion in $\Lambda$. \\
\indent In this paper, we employ the GALE (Global Acoustic Linearized Euler) code \citep{2020arXiv201103131S}, simulating the propagation of acoustic waves through shallow and deep single-cell, as well as strong and weak double-cell regimes of meridional background flows---generated by the mean-field non-linear hydrodynamic and dynamo models of \citet{2018ApJ...854...67P,2019ApJ...887..215P}. The stochastic excitation of oscillation sources over background flows generated by these models allows us to perform a systematic examination of realization noise in the helioseismic signatures generated by each regime. This investigation offers a baseline for the low-end of variance in travel-time measurements that characterize single-cell and double-cell meridional circulation---resulting from minimal parameter changes near the base of the tachocline. Previous conclusions on the nature of meridional circulation structure and the bounds of realization noise have been inferred with the aid of ad-hoc models \citep{2013ApJ...774L..29Z,2013ApJ...762..132H,2015ApJ...813..114R,2019_chen,2020Sci...368.1469G}; we analyze physics-based meridional velocity profiles in order to help constrain the large variance of possible internal structures. We show that a definitive statement on the nature of single- or double-cell meridional circulation may not be possible even if the problems with CtoL corrections are fully resolved, as the difference between the two regimes may fall within the variance of the realization noise.\\
\indent This paper is organized as follows. In \S\ref{sec:model} we briefly describe the formulation of the GALE code. In \S\ref{sec:method} we outline the deep focusing method used to measure travel-time differences of acoustic rays propagating throughout the solar interior, and in \S\ref{sec:MM}, we present the results of these measurements on four models (M1, M2, K1 \& K2) of meridional circulation. In \S\ref{sec:conclusions} we offer an analysis and discussion of these results.

\section{Model Background} \label{sec:model}

The GALE (Global Acoustic Linearized Euler) algorithm \citep{2020arXiv201103131S} solves the conservation form of the linearized compressible Euler equations on a fully global 3-dimensional grid: $0 \le \phi \le 2\pi$, $0 < \theta < \pi$, $0 < r \le R_{\odot}$. The linear approximation is solved for perturbations of the potential flow field (denoted by a prime), over background field terms (denoted by a tilde) derived from the standard solar model S \citep{1996Sci...272.1286C}.

\begin{equation}\label{eq:gov1}
  \dfrac{\partial \rho'}{\partial t} + \Upsilon' = 0 \ ,
\end{equation}
\begin{equation}\label{eq:gov2}
  \dfrac{\partial\Upsilon'}{\partial t} + \boldsymbol{\nabla}:\left(\mathbf{m}'\tilde{\mathbf{u}} + \tilde{\rho}\tilde{\mathbf{u}}\mathbf{u}'\right) = -\nabla^{2}\left(p'\right) - \nabla\cdot\left(\rho'\tilde{g}_{r}\mathbf{\hat{r}}\right) + \nabla\cdot S\mathbf{\hat{r}} \ ,
\end{equation}
\begin{equation}\label{eq:gov3}
  \dfrac{\partial p'}{\partial t} = - \dfrac{\Gamma_{1}\tilde{p}}{\tilde{\rho}}\left(\nabla\cdot\tilde{\rho}\mathbf{u}'+ \rho'\nabla\cdot\tilde{\mathbf{u}} - 
  \dfrac{p'}{\tilde{p}}\tilde{\mathbf{u}}\cdot\nabla\tilde{\rho} + \tilde{\rho} \mathbf{u}'\cdot\dfrac{N^{2}}{g}\mathbf{\hat{r}}\right) \ .
\end{equation}

\noindent $\Upsilon$ is defined as the divergence of the momentum field $\mathbf{m}$ ($\Upsilon = \nabla\cdot\mathbf{m}= \nabla\cdot\rho\mathbf{u}$), computing perturbations in the potential flow field and omitting solenoidal terms in our governing equations. Perturbations are initiated by a randomized source function ($S$), mimicking the stochastic excitation of acoustic modes in the convective interior \citep{1984PhDT........34W}. Governing Eqns. (\ref{eq:gov1}) - (\ref{eq:gov3}) are computed in the pseudo-spectral regime through spherical harmonic decomposition ($f = \sum_{lm} a Y_{lm}$) of field terms ($\Upsilon$, $\rho$, $p$, $\mathbf{u}$) using the Libsharp spherical harmonic library \citep{2013A&A...554A.112R}. The current form of our governing equations (Eqns. \ref{eq:gov1} - \ref{eq:gov3}) are solved in a vector spherical harmonic (VSH) basis, while the material derivative is solved in its Cauchy conservation form ($\boldsymbol{\nabla}:\left(\mathbf{m}'\tilde{\mathbf{u}} + \tilde{\rho}\tilde{\mathbf{u}}\mathbf{u}'\right)$) using a symmetric tensor spherical harmonic (TSH) basis. This formulation allows for the use of recursion relations to compute derivatives tangent to the surface of the sphere (${\partial}/{\partial\theta},{\partial}/{\partial\phi}$), resulting in the 1D radial relations we solve in our time-discretization scheme. \\
\indent The conservation of energy (Eq. \ref{eq:gov3}) is computed as a relationship between pressure and the compressible momentum field in the adiabatic approximation \citep{christensen14}, where heat transfer is neglected. Convective action in the background model is treated separately by isolating the Brunt-V\"ais\"al\"a frequency ($N^{2}$). In order to avoid growing convective instabilities, we set the slightly negative values of the Brunt-V\"ais\"al\"a frequency in the convection zone ($N^{2} < 0$) to zero---obviating the need to modify background profiles of density, pressure and the adiabatic ratio ($\Gamma_{1}$), see \citet{2006ApJ...648.1268H,2007ApJ...666..547P,2014SoPh..289.1919P}. A full description and validation of the model's algorithm can be found in \citet{2020arXiv201103131S}.

\section{Method}\label{sec:method}

\indent The spatial resolution of the model is set by the spherical harmonic degree $l=200$. This value corresponds to an azimuthal resolution of $N_{\phi} = 600$ and a latitudinal resolution of $N_{\theta} = 450$, chosen to avoid aliasing errors during sampling. This resolution allows for the detection of signals throughout the convection zone to an upper limit of $\sim 0.96R_{\odot}$. The temporal cadence of the model is 3 seconds with data being saved at 1 minute intervals for a total of $~67$ hours ($4000$ minutes)---a time-scale too short to effectively resolve signals in the convection zone \citep{2008ApJ...689L.161B}. We can improve the signal-to-noise ratio (SNR) by increasing the background velocities in our model by a factor of 36 (see \citet{2013ApJ...762..132H}), mimicking the SNR we would expect in approximately a decade of solar observations---within the operational time-frame of HMI.\\
\indent Acoustic oscillations traveling through the solar interior are advected by regimes of mean mass flows in the convection zone. Measuring perturbations on the solar surface allows for the application of local helioseismology techniques (see \citet{2005LRSP....2....6G}) in order to infer the structure of these internal flow velocities. We employ the technique of deep focusing \citep{2005LRSP....2....2C,2009ApJ...702.1150Z}, in which radial velocity perturbtaions at two points on the model surface, separated by some angular distance ($\Delta$), are cross-correlated. The resulting correlation signal forms the charectristic profile of a Gabor wavelet \citep{1997ASSL..225..241K,1999_giles,2007ApJ...659.1736N}, which can be described by the function:

\begin{equation}\label{eq:gabor}
\Psi(\tau,\Delta) \propto \sum\limits_{\delta v} \cos\left[\omega_{0}\left(\tau-\dfrac{\Delta}{v} \right)\right]\exp\left[-\dfrac{\delta\omega^{2}}{4}\left(\tau-\dfrac{\Delta}{u} \right)^{2}\right] \ .
\end{equation}

\noindent $\tau$ is the time-lag between the signals, $\omega_{0}$ and $\delta\omega$ are the central frequency and width of the frequency filter respectively. $v = \omega/\sqrt{l(l+1)}$ is the horizontal phase velocity and $u = \partial\omega/\partial k_{h}$ is the horizontal group velocity. The function is summed over a narrow range of phase velocities $\delta v$. We fit this function (Eq. \ref{eq:gabor}) to our cross-correlated signal using the iterative Levenberg-Marquardt method, providing an estimate of the time-lag from the group and phase velocities.\\
\indent In order to measure the impact of meridional flow fields on acoustic rays, we subtract the time-lag ($\tau$) of southward traveling waves from their northward traveling counterparts ($\delta\tau_{NS}$). The ray-path approximation \citep{1999_giles} then offers a basis for inferring the magnitude of the velocity field in the direction of the travelling acoustic ray (assuming the path of the ray $\Gamma_{0}$ remains unperturbed). 

\begin{equation}\label{eq:raypath}
  \delta\tau = -2\int_{\Gamma_{0}} \dfrac{\mathbf{u}\cdot\mathbf{n}}{c^{2}} ds \ .
\end{equation}

\begin{figure}[!htb]
\center
\includegraphics[width=13cm]{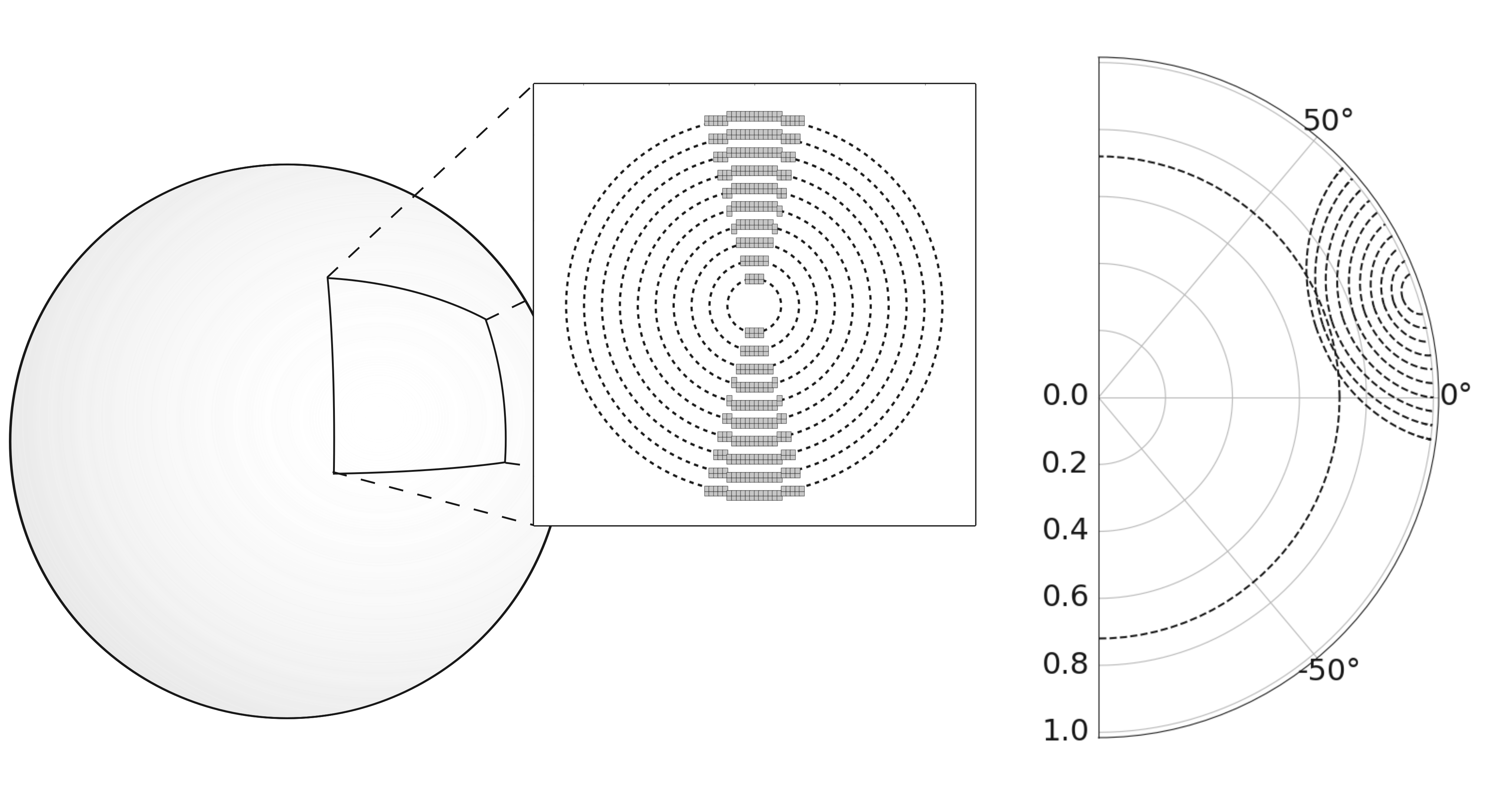}
\caption{An illustration of the deep-focusing method. Left) A $60^{\circ}\text{ x }60^{\circ}$ patch is remapped into azimuthal equidistant coordinates, pixels are selected in $30^{\circ}$ wide northern and southern sectors (2 pixels in radius). Right) The acoustic ray-paths associated with the selected travel distances.}
\label{fig:deepfoc}
\end{figure}

\noindent Although this functions omits non-linear effects of large flow velocities, it serves as a basic approximation of travel-time differences---seen by computed travel-time comparisons in Fig. \ref{fig:TD_err_F_rm}.\\
\indent This method is applied to the surface mesh-grid of our model, sampled at approximately $300\text{ km}$ above the photosphere ($1.0004 R_{\odot}$). Each pixel ($N_{\theta}\text{ x }N_{\phi} = 450\text{ x }600$) is treated as a center point around which cubic hermite splines are used to remap a $60^{\circ}\text{ x }60^{\circ}$ patch into azimuthal equidistant coordinates (Postel's proejection) at a resolution of approximately $0.6^{\circ}$ per pixel. Great circle distances ($\Delta$) are selected at every interval ($1.2^\circ$) from $0^{\circ}$ to $60^{\circ}$, for which $30^{\circ}$ wide sectors in the north and south (2 pixels in radius) are averaged and cross-correlated with each other. This signal can be enhanced by averaging over every point in the longitude ($N_{\phi} = 600$) and smoothed by averaging $\pm 3^{\circ}$ ($\pm 5\text{ pix}$) in the latitude. We can further smooth our data by averaging the diameter of each great circle over $\pm 2.4^{\circ}$, where travel time offsets are interpolated to an estimated time-lag using the ray-path approximation (Thank you to Dr. Ruizhu Chen for suggesting this smoothing technique). The application of this method allows for the measurement of acoustic travel times and inferences of meridional velocities throughout the convection zone---from the convective-radiative interface at the tachocline ($0.70 R_{\odot}$) to near the solar surface ($0.96 R_{\odot}$). An illustration of the remapping and pixel selection technique, along with corresponding acoustic ray-paths through the model interior can be seen in Fig. \ref{fig:deepfoc}.

\section{Meridional Profiles}\label{sec:MM}

\begin{figure}[]
\center
\includegraphics[width=17cm]{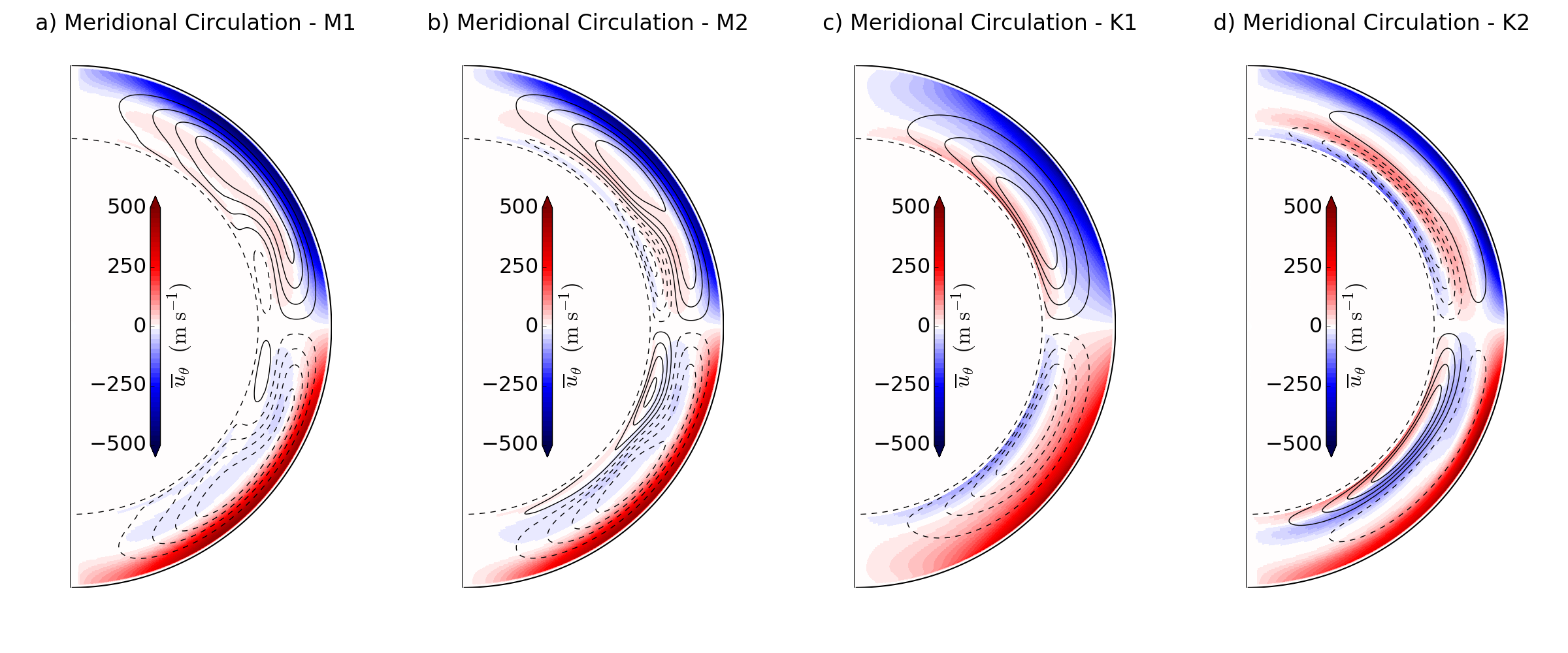}
\caption{Latitudinal velocities ($\tilde{u}_{\theta}$), generated by the mean-field models of \citet{2019ApJ...887..215P} (M1 and M2) and \citet{2018ApJ...854...67P} (K1 and K2---referred to as M2 and M3 in their paper). a) Single-cell meridional circulation with a shallow return flow at $\sim 0.80R_{\odot}$. b) Double-cell meridional circulation with weak reversal. c) Single-cell meridional circulation with a deep return flow near the base of the tachocline. d) Double-cell meridional circulation with strong reversal. Solid and dashed contours represent counterclockwise and clockwise rotation respectively. Meridional circulation models are amplified by a factor of 36 (see Section \ref{sec:method}).}
\label{fig:MCv}
\end{figure}

\indent We apply the deep focusing method (\S\ref{sec:method}) to four profiles of meridional circulation generated by non-linear mean-field hydrodynamic and dynamo models. The first two are a shallow single-cell, with return flow at approximately $0.80R_{\odot}$, and a double-cell meridional circulation profile referred to as M1 and M2 respectively in \citet{2019ApJ...887..215P}. The next two are a single-cell meridional circulation model with a deep return flow situated near the base of the tachocline---based on the mean-field model of \citet{2004ARep...48..153K}, as well as a double-cell meridional circulation profile with a stronger return flow induced by gyroscopic pumping in the same model. They are described as models M2 and M3 respectively in \citet{2018ApJ...854...67P}; in this paper we will refer to them model as K1 and K2 in order to avoid confusion. These meridional velocities are used as the background terms ($\tilde{u}_{r}$,$\tilde{u}_{\theta}$) in our governing equations (\ref{eq:gov1}-\ref{eq:gov3}). The latitudinal velocities ($\tilde{u}_{\theta}$) for the models (M1, M2, K1 \& K2) are shown in Fig. \ref{fig:MCv}, with streamlines representing the circulation profile.

\begin{figure}[]
\center
\includegraphics[width=15cm]{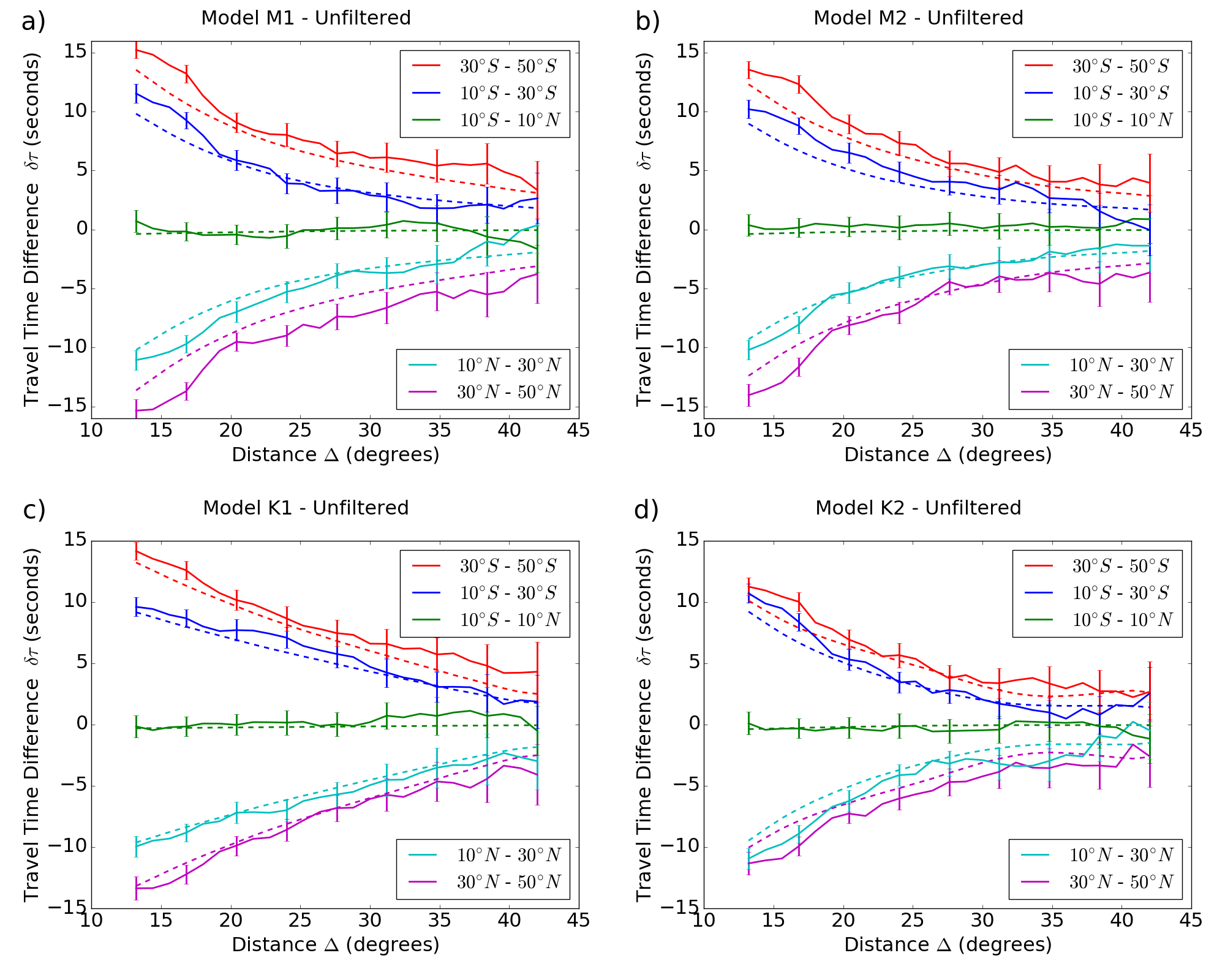}
\caption{The N-S travel-time differences ($\delta\tau_{NS}$) as a function of travel distance ($\Delta$) for models M1 (a), M2 (b), K1 (c) and K2 (d). The travel-time measurements are shown unfiltered for 5 latitudinal averages spanning $30^{\circ}\text{N} - 50^{\circ}\text{N}$, $10^{\circ}\text{N} - 30^{\circ}\text{N}$, $10^{\circ}\text{S} - 10^{\circ}\text{N}$, $10^{\circ}\text{S} - 30^{\circ}\text{S}$, $30^{\circ}\text{S} - 50^{\circ}\text{S}$. Dashed lines are theoretical times computed using the ray-path approximation (Eq. \ref{eq:raypath}). Error bars show the standard deviation of the measured realization noise ($\sigma_{NS}$, Eq. \ref{eq:error}). Meridional circulation models are amplified by a factor of 36 (see Section \ref{sec:method}).}

\label{fig:TD_err_U}
\end{figure}

\begin{figure}[]
\center
\includegraphics[width=15cm]{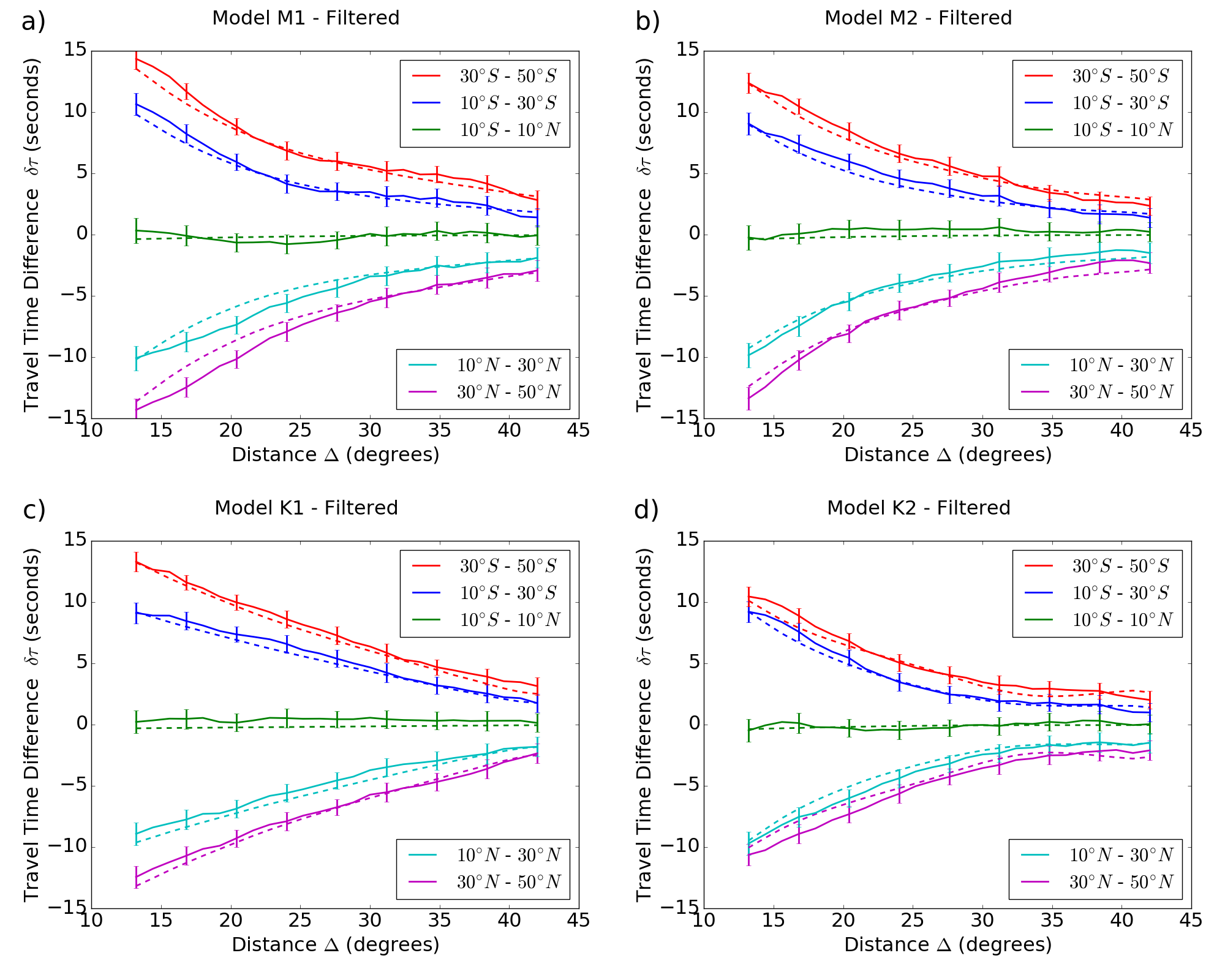}
\caption{The N-S travel-time differences ($\delta\tau_{NS}$) as a function of travel distance ($\Delta$) for models M1 (a), M2 (b), K1 (c) and K2 (d). The travel-time measurements are shown under the application of a Gaussian phase-speed filter ($\sigma = 0.05v_{p}$) for 5 latitudinal averages spanning $30^{\circ}\text{N} - 50^{\circ}\text{N}$, $10^{\circ}\text{N} - 30^{\circ}\text{N}$, $10^{\circ}\text{S} - 10^{\circ}\text{N}$, $10^{\circ}\text{S} - 30^{\circ}\text{S}$, $30^{\circ}\text{S} - 50^{\circ}\text{S}$. Dashed lines are theoretical times computed using the ray-path approximation (Eq. \ref{eq:raypath}). Error bars show the standard deviation of the measured realization noise ($\sigma_{NS}$, Eq. \ref{eq:error}). Meridional circulation models are amplified by a factor of 36 (see Section \ref{sec:method}).
}
\label{fig:TD_err_F}
\end{figure}

\indent We characterize internal meridional flow profiles by plotting N-S travel-time differences ($\delta\tau_{NS}$) as a function of their travel distance ($\Delta =  12^{\circ} - 42^{\circ}$)---corresponding to an increasing depth in the convection zone ($r = 0.93R_{\odot} - 0.72R_{\odot}$). In Fig. \ref{fig:TD_err_U}, we show 5 latitudinal averages ($30^{\circ}\text{N} - 50^{\circ}\text{N}$, $10^{\circ}\text{N} - 30^{\circ}\text{N}$, $10^{\circ}\text{S} - 10^{\circ}\text{N}$, $10^{\circ}\text{S} - 30^{\circ}\text{S}$, $30^{\circ}\text{S} - 50^{\circ}\text{S}$) of these travel-time differences for our regimes of meridional circulation. Measured travel-time differences (solid lines) are compared to theoretical travel-time differences (dashed lines) computed using the ray-path approximation (Eq. \ref{eq:raypath}). Randomized functions are used to generate the oscillatory signal in our source ($S$, Eq. \ref{eq:gov2}, see \citet{2020arXiv201103131S} for a detailed description) resulting in unique profiles of realization noise---of which four different profiles are shown for models M1, M2, K1 and K2 (Fig. \ref{fig:TD_err_U}). To measure the bounds of our noise, we use 100 unique source functions on a model with no background flows; the noise can then be characterized as the standard deviation ($\sigma_{NS}$) of the measured signal from zero:

\begin{equation}\label{eq:error}
  \sigma_{NS} = \sqrt{\dfrac{1}{N}\sum_{i=1}^{N}  \delta\tau_{i}^{2}} \ .
\end{equation}

\noindent Error-bars in Fig. \ref{fig:TD_err_U} show one standard deviation of the noise. We can improve the SNR in our measurements significantly---especially at larger travel distances, by applying a phase-velocity filter to our data prior to deep focusing. The filter is defined as a Gaussian function in phase-velocity space \citep{2007ApJ...659.1736N} with a width of $\sigma=0.05 v$, where $v$ is the phase speed ($\omega/l(l+1)$). The resulting travel-time differences can be seen in Fig. \ref{fig:TD_err_F}.

\begin{figure}[]
\center
\includegraphics[width=15cm]{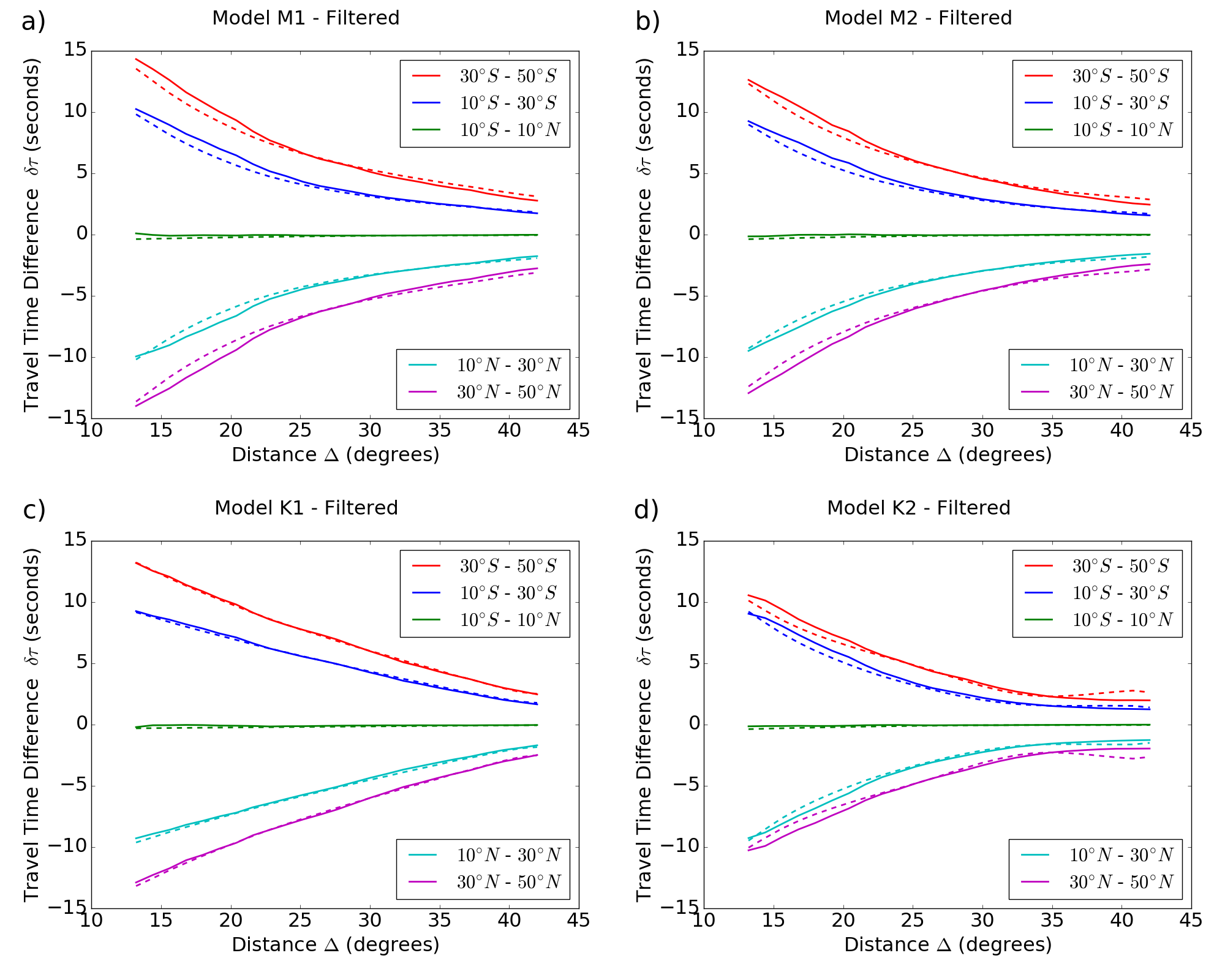}
\caption{The N-S travel-time differences ($\delta\tau_{NS}$) as a function of travel distance ($\Delta$) for models M1 (a), M2 (b), K1 (c) and K2 (d) under the application of Gaussian phase-speed filter ($\sigma = 0.05v_{p}$) with profiles of noise subtracted, showing 5 latitudinal averages spanning $30^{\circ}\text{N} - 50^{\circ}\text{N}$, $10^{\circ}\text{N} - 30^{\circ}\text{N}$, $10^{\circ}\text{S} - 10^{\circ}\text{N}$, $10^{\circ}\text{S} - 30^{\circ}\text{S}$, $30^{\circ}\text{S} - 50^{\circ}\text{S}$. Dashed lines are theoretical times computed using the ray-path approximation (Eq. \ref{eq:raypath}). Meridional circulation models are amplified by a factor of 36 (see Section \ref{sec:method}).}
\label{fig:TD_err_F_rm}
\end{figure}

\indent Comparing the signals from the four models, we see noticeable variations between three of them. As we move from a deep single-cell profile (K1) to the shallow single-cell and weak-reversal double-cell profiles (M1 \& M2), further onto the strong-reversal double-cell profile (K2) We see a trend of increasing curvature showing a more rapid decrease in travel-time differences with travel distance. These trends result in gaps wider than one standard deviation of the noise throughout a large part of the convective interior. The shallow single-cell and double-cell regimes (M1 and M2 respectively) however, fall within one standard deviation of the range of realization noise, even with the significant increase in the SNR through the application of a phase-velocity filter. These results have positive implications for our ability to distinguish between deep and shallow single-cell meridional circulation, as well as profiles of double-cell circulation with strong reversals. Unfortunately, the differences between shallow single- and double-cell profiles are much more subtle; using current helioseismology techniques, within the time-frame of HMI measurements ($\sim$ 10 years), any definitive statements on whether meridional circulation has one or two cells may be difficult to make.\\
\indent In order to highlight systematic errors present in our model, as well as compare our measurements to ray-path theory, we subtract computed travel-time differences of models with no background flows from the travel-time differences of the models with meridional circulation profiles that were excited with the same source profile ($S$, see Eq. \ref{eq:gov2}). This results in realization noise being removed from travel-time differences in models M1, M2, K1 and K2, for the phase-speed filtered data-set (Fig. \ref{fig:TD_err_F_rm})---revealing a close and consistent agreement of the ray-path approximation with the travel-time differences in the K1 model. As travel-time difference profiles begin to fall off more steeply with travel distance however, our results begin to diverge slightly from the approximation, culminating in an inability to match the travel-time difference increase seen at large depths in model K2. Whether this is due to intrinsic systematic errors in our model or in the approximation itself deserves an in-depth investigation that is outside the scope of this paper.

\begin{figure}[]
\center
\includegraphics[width=7.5cm]{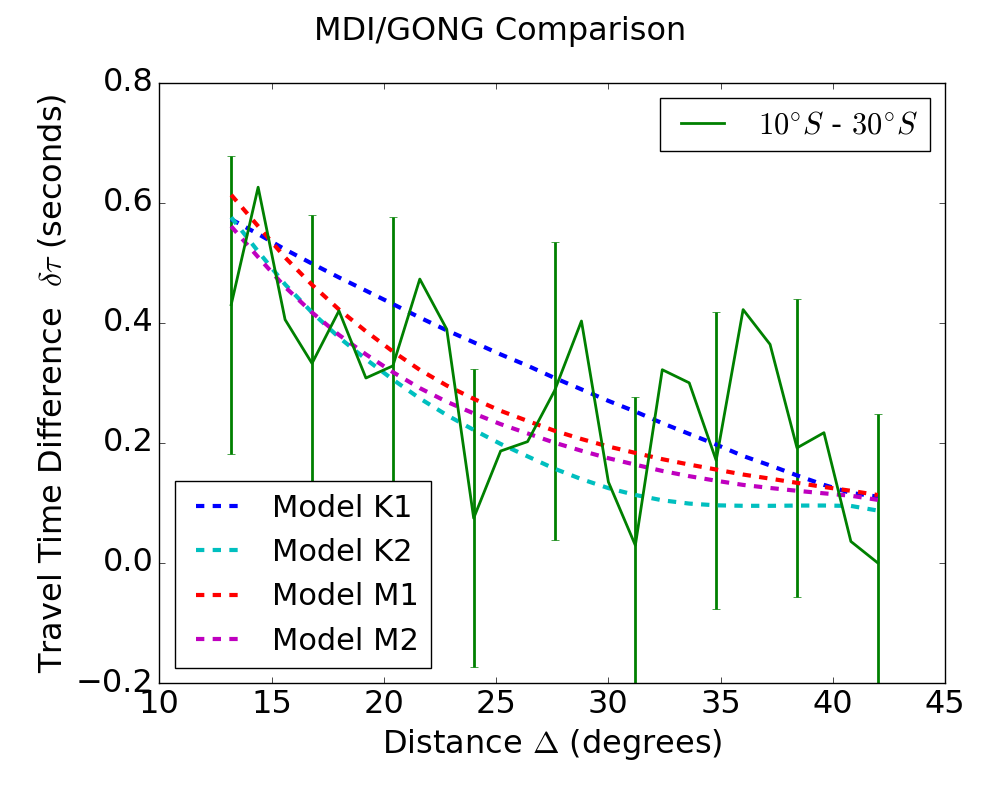}
\caption{The N-S travel-time differences ($\delta\tau_{NS}$) as a function of travel distance ($\Delta$) for MDI/GONG data published by \citet{2020Sci...368.1469G}. Latitude ranges in both hemispheres ($10^{\circ}\text{N} - 30^{\circ}\text{N}$, $10^{\circ}\text{S} - 30^{\circ}\text{S}$) are averaged in order to reduce noise and are compared to dashed lines representing latitudinal averages for models K1, K2, M1 and M2 as measured using the ray-path approximation (Eq. \ref{eq:raypath}). Error-bars are computed as the standard deviation ($\sigma$, Eq. \ref{eq:error}) of the travel-time differences in the $10^{\circ}\text{N} - 10^{\circ}\text{S}$ latitude range from zero.}
\label{fig:TD_cmp}
\end{figure}

\indent We further compare our models to the publicly available data from the analysis of \citet{2020Sci...368.1469G} (Fig. \ref{fig:TD_cmp}), scaling the travel-time differences computed using the ray-path approximation in our models (M1, M2, K1 \& K2), to their travel-time measurements using MDI/GONG observations for Solar Cycle 24 (2008-2019). It is apparent that the level of noise is too large to draw significant conclusions distinguishing single- and double-cell regimes of meridional circulation.

\newpage

\section{Discussion}\label{sec:conclusions}

\indent The question on the nature of meridional circulation remains an outstanding one. Large variations in inferences have been made using similar local helioseismology techniques---from a double-cell meridional circulation profile seen in HMI observations \citep{2013ApJ...774L..29Z, 2019_chen} to recent re-assertions of a single cell profile from MDI/GONG data \citep{2020Sci...368.1469G}. The feasibility of the two regimes has been explored in recent years, with convectively driven MHD and HD models reproducing the necessary environment for both.\\
\indent The actual differences may lie somewhere in between the large variations inferred in the two models. To that end we test the feasibly of discerning between meridional profiles with a deep and shallow single-cell as well as strong and weak secondary-cell profiles, generated by the same mean-field model, operating with minimal parameter changes (K1 \& K2 in \citet{2018ApJ...854...67P}---referred to as M2 and M3 in their paper; M1 \& M2 in \citet{2019ApJ...887..215P}). Using the GALE code \citep{2020arXiv201103131S} to simulate the stochastic excitation of acoustic modes throughout the convective interior, we apply the deep focusing method to measure the resulting realization noise in travel-time differences. Noise is one of the main factors that we have to contend with when inferring structures from observations. Even if the systematic CtoL effect is fully resolved, we are bound by the time-frame of measurements that we currently have. The high levels of noise in these measurements leave large uncertainties that may be the source of the divergent conclusions made on the nature of meridional circulation. Physics-based models can be used to constrain results in the context of a broader complex system. We analyze differences between a shallow single-cell and a weak-reversal double-cell regime, generated by a mean-field simulation that uses a physics-based model of gyroscopic pumping to induce the reverse-flow cell near the base of the tachocline. In these models we see that even with the formation of a reverse flow, the travel-time differences may fall well within one standard deviation of error for both phase-speed filtered and unfiltered deep-focusing measurements. These models (M1 \& M2) provide the low-end of the baseline for variance between the two regimes, however, we can also examine physics-based profiles that may be more consistent with single-cell \citep{2020Sci...368.1469G} inferences such as K1 or double-cell \citep{2013ApJ...774L..29Z, 2019_chen} inferences such as K2. The deep single-cell profile (K1) and strong-reversal double-cell profile (K2) show large enough differences that a distinction with a relatively high degree of confidence is feasible. For now, however, the best way to constrain these solutions is by looking at the broader impacts of these profiles within the context of their effects on global solar dynamics \citep[see][]{2013IAUS..294..399K}. In \citet{2018ApJ...854...67P} we see that an unavoidable effect of increasing the strength of the return flow is a profile of differential rotation that is inconsistent with inferences of global helioseismology \citep{2011JPhCS.271a2061H}. While mean-field modeling does not fully capture the structure of the solar interior, a consistency with our current theoretical understanding of the system can lend confidence to our results. While meaningful conclusions on the nature of the return flow are within the reach of local helioseismology techniques, we see that currently, mean-field model-based profiles of meridional circulation show that any definitive statements on the prevalence of either a single- or double-cell regime being made should be taken with a grain of salt.

\acknowledgments

\indent AMS would like to thank the heliophysics modeling and simulation team at NASA Ames Research Center for help and guidance. VVP would like to thank the FR project II.16.3 of ISTP SB RAS for their support. This work is supported by the NASA grants: 80NSSC19K0630, 80NSSC19K1436, NNX14AB70G, NNX17AE76A, 80NSSC20K0602


\begin{thebibliography}{}


\bibitem[Bekki \& Yokoyama(2017)]{2017ApJ...835....9B} Bekki, Y., Yokoyama, T.\ 2017, \apj, 835, 9

\bibitem[Bogart et al.(2015)]{2015ApJ...807..125B} Bogart, R.~S., Baldner, C.~S., Basu, S.\ 2015, \apj, 807, 125

\bibitem[B\"oning et al.(2017)]{2017ApJ...845....2B}  B\"oning, V., Roth, M., Jackiewicz, J., Kholikov, S.\ 2017, \apj, 845, 2

\bibitem[B\"oning(2018)]{2018IAUS..340...13B}  B\"oning, V.\ 2018, IAUS, 340, 13


\bibitem[Braun \& Birch(2008)]{2008ApJ...689L.161B} Braun, D.~C., Birch, A.~C.\ 2008, \apj, 689, 161
    

\bibitem[Brun \& Toomre(2002)]{2002ApJ...570..865B} Brun A.~S., Toomre J.\ 2002, \apj, 570, 865


\bibitem[Chaplin et al.(1996)]{1996MNRAS.283L..31C} Chaplin, W.~J., Elsworth, Y., Isaak, G.~R., et al.\ 1996, MNRAS, 283, 31


\bibitem[Charbonneau(2005)]{2005LRSP....2....2C} Charbonneau, P.\ 2005, LRSP, 2, 2




\bibitem[Chen \& Zhao(2017)]{2017ApJ...849..144C} Chen, R., Zhao. J.\ 2017, \apj, 849, 144

\bibitem[Chen \& Zhao(2018)]{2018ApJ...853..161C} Chen, R., Zhao. J.\ 2018, \apj, 853, 161


\bibitem[Chen(2019)]{2019_chen} Chen, R.\ 2019, Ph.D. Thesis, Stanford University


\bibitem[Christensen-Dalsgaard et al.(1996)]{1996Sci...272.1286C} Christensen-Dalsgaard, J., Dappen, W., Ajukov, S.~V.\ 1996, Sci, 272, 1286 
 
\bibitem[Christensen-Dalsgaard(2002)]{2002RvMP...74.1073C}  Christensen-Dalsgaard, J.,\ 2002, RvMP, 74, 1073 

\bibitem[Christensen-Dalsgaard(2014)]{christensen14}  Christensen-Dalsgaard, J.,\ 2014, D.f.I.Print, Aarhus.


\bibitem[Domingo et al.(1995)]{1995SoPh..162....1D} Domingo, V., Fleck, B., Poland, A.~I.\ 1995, SoPh, 162, 1


\bibitem[Duvall(1979)]{1979SoPh...63....3D} Duvall, T.~L., Jr.\ 1979, SoPh, 63, 3


\bibitem[Featherstone \& Miesch(2015)]{2015ApJ...804...67F} Featherstone, N.~A., Miesch M.~S.\ 2015, \apj, 804, 67
	

\bibitem[Gastine et al.(2014)]{2014MNRAS.438L..76G} Gastine, T., Yadav, R.~K., Morin, J., Reiners, A., Wicht, J.\ 2014, MNRAS, 438, 76
 

\bibitem[Giles et al.(1997)]{1997Natur.390...52G} Giles, P.~M., Duvall, T.~L., Scherrer, P.~H., Bogart, R.~S.\ 1997, Nature, 390, 52


\bibitem[Giles(1999)]{1999_giles} Giles, P.~M.\ 1999, Ph.D. Thesis, Stanford University



\bibitem[Gizon \& Birch(2004)]{2004ApJ...614..472G} Gizon, L, Birch, A.~C.\ 2004, \apj, 614, 472
 

\bibitem[Gizon \& Birch(2005)]{2005LRSP....2....6G} Gizon, L, Birch, A.~C.\ 2005, LRSP, 2, 6
 


\bibitem[Gizon et al.(2020)]{2020Sci...368.1469G} Gizon, L., Cameron, R.~H., Pourabdian, M., et al.\ 2020, Sci, 368, 1469


\bibitem[Guerrero et al.(2013)]{2013ApJ...779..176G} Guerrero, G., Smolarkiewicz, P.~K., Kosovichev, A.~G., Mansour N.~N.\ 2013, \apj, 779, 176


\bibitem[Hanasoge et al.(2006)]{2006ApJ...648.1268H}  Hanasoge, S.~M., Larsen, R.`M., Duvall, T.~L., Jr., et al.\ 2006, \apj, 648, 1268 


\bibitem[Hartlep et al.(2013)]{2013ApJ...762..132H}  Hartlep, T., Zhao, J., Kosovichev, A.~G., Mansour, N.~N.\ 2013, \apj, 762, 132


\bibitem[Harvey et al.(1996)]{1996Sci...272.1284H}  Harvey, J.~W., Hill, F., Hubbard, R.~P., et al.\ 1996, Sci, 272, 1284

  
\bibitem[Hathaway(1996)]{1996ApJ...460.1027H} Hathaway, D.~H.\ 1996, \apj, 460, 1027
    

\bibitem[Hathaway et al.(2003)]{2003ApJ...589..665H} Hathaway, D.~H., Nandy, D., Wilson, R.~M., Reichmann, E.~J.\ 1996, \apj, 589, 665
    

\bibitem[Hathaway(2012)]{2012ApJ...760...84H} Hathaway, D.~H.\ 2012, \apj, 760, 84



\bibitem[Hotta \& Yokoyama(2011)]{2011ApJ...740...12H} Hotta, H., Yokoyama, T.\ 2011, \apj, 740, 12
	

\bibitem[Howe et al.(2011)]{2011JPhCS.271a2061H} Howe, R., Larson, T.~P., Schou, J., et al.\ 2011, JPhCS, 271, 2061

	
\bibitem[Jackiewicz et al.(2015)]{2015ApJ...805..133J} Jackiewicz, J., Serebryanskiy, A., Kholikov, S.\ 2015, \apj, 805, 133

 

\bibitem[K\"apyl\"a et al.(2014)]{2014A&A...570A..43K}  K\"apyl\"a, P.~J., K\"apyl\"a, M.~J., Brandenburg, A.\ 2014, \aap, 570, 43


\bibitem[Kichatinov \& Rudiger(1993)]{1993A&A...276...96K} Kichatinov, L.~L., Rudiger, G.\ 1993, \aap, 276, 96

\bibitem[Kichatinov(2004)]{2004ARep...48..153K} Kichatinov, L.~L.\ 2004, ARep, 48, 153


\bibitem[Kitchatinov \& Olemskoy(2012)]{2012MNRAS.423.3344K}  Kitchatinov, L.~L., Olemskoy, S.~V.\ 2012, MNRAS, 423, 3344

    
\bibitem[Kitchatinov(2013)]{2013IAUS..294..399K} Kitchatinov, L.~L.\ 2013, IAUS, 294, 399
	
	
\bibitem[Kholikov et al.(2014)]{2014ApJ...784..145K} Kholikov, S., Serebryanskiy, A., Jackiewicz, J.\ 2014, \apj, 784, 145
	
    
\bibitem[Komm et al.(2015)]{2015SoPh..290.3113K} Komm, R., González Hernández, I., Howe, R., Hill, F.\ 2015, SoPh, 290, 3113
	
	
\bibitem[Kosovichev \& Duvall(1997)]{1997ASSL..225..241K} Kosovichev, A.~G., Duvall, T.~L., Jr.\ 1997, ASSL, 225, 241
	

\bibitem[K\"uker et al.(2011)]{2011A&A...530A..48K}  K\"uker, M., Rüdiger, G., Kitchatinov, L.~L.\ 2011, \aap, 530, 48

	
\bibitem[Liang \& Chou(2015)]{2015ApJ...805..165L} Liang, Z-C., Chou, D-Y.\ 2015, \apj, 805, 165
    

\bibitem[Lin \& Chou(2018)]{2018ApJ...860...48L} Lin, C-H., Chou, D-Y.\ 2018, \apj, 860, 48
    

\bibitem[Matilsky et al.(2019)]{2019ApJ...871..217M} Matilsky, L.~I.; Hindman, B.~W., Toomre, J.\ 2019, \apj, 871, 217


\bibitem[Miesch et al.(2006)]{2006ApJ...641..618M} Miesch, M.~S., Brun A.~S., Toomre J.\ 2006, \apj, 641, 618

\bibitem[Mitra-Kraev \& Thompson(2007)]{2007AN....328.1009M}  Mitra-Kraev, U., Thompson, M.~J.\ 2007, Astron.Nachr., 328, 1009


\bibitem[Nigam et al.(2007)]{2007ApJ...659.1736N} Nigam, R., Kosovichev, A.~G., Scherrer, P.~H.\ 2007, \apj, 659, 1736
	

\bibitem[Papini et al.(2014)]{2014SoPh..289.1919P}  Papini, E.; Gizon, L.; Birch, A. C.\ 2014, SoPh, 289, 1919


\bibitem[Parchevsky \& Kosovichev(2007)]{2007ApJ...666..547P} Parchevsky, K.~V., Kosovichev, A.~G.\ 2007, \apj, 666, 547 


\bibitem[Pesnell et al.(2012)]{2012SoPh..275....3P} Pesnell, W.~D., Thompson, B.~J., Chamberlin, P.~C.\ 2012, SoPh, 275, 3
    


\bibitem[Pipin \& Kosovichev(2018)]{2018ApJ...854...67P} Pipin, V.~V., Kosovichev, A.~G.\ 2018, \apj, 854, 67
     
    
\bibitem[Pipin \& Kosovichev(2019)]{2019ApJ...887..215P} Pipin, V.~V., Kosovichev, A.~G.\ 2019, \apj, 887, 215
    

\bibitem[Rajaguru \& Antia(2015)]{2015ApJ...813..114R} Rajaguru, S.~P., Antia, H.~M.\ 2015, \apj, 813, 114 


\bibitem[Rajaguru \& Antia(2020)]{2020arXiv200412708R} Rajaguru, S.~P., Antia, H.~M.\ 2020, arXiv:2004.12708 


\bibitem[Reinecke \& Seljebotn(2013)]{2013A&A...554A.112R}  Reinecke, M., Seljebotn, D.~S.\ 2013, \aap, 554, 112 


\bibitem[Rempel(2005)]{2005ApJ...622.1320R} Rempel, M.\ 2005, \apj, 622, 1320

 
\bibitem[Ruediger(1989)]{1989drsc.book.....R} Ruediger, G.\ 1989, Akademie Verlag, Berlin


\bibitem[Scherrer et al.(1995)]{1995SoPh..162..129S} Scherrer, P.~H., Bogart, R.~S., Bush, R.~I., et al.\ 1995, SoPh, 162, 129

\bibitem[Scherrer et al.(2012)]{2012SoPh..275..207S} Scherrer, P.~H., Schou, J., Bush, R.~I., et al.\ 2012, SoPh, 275, 207

\bibitem[Stejko et al.(2021)]{2020arXiv201103131S} Stejko, A.~S., Kosovichev, A.G., Mansour, N.~N.\ 2021, \apjs, 253, 9


\bibitem[Schad et al.(2013)]{2013ApJ...778L..38S} Schad, A., Timmer, J., Roth, M.\ 2013, \apj, 778, 38

    

\bibitem[Ulrich(2010)]{2010ApJ...725..658U} Ulrich, R.~K.\, \apj, 725, 658


\bibitem[Woodard(1984)]{1984PhDT........34W} Woodard, M.~F.\ 1984, Ph.D. Thesis,  University of California, San Diego 


\bibitem[Zhao \& Kosovichev(2004)]{2004ApJ...603..776Z} Zhao, J., Kosovichev, A.~G.,\ 2004, \apj, 603, 776


\bibitem[Zhao et al.(2009)]{2009ApJ...702.1150Z} Zhao, J., Hartlep, T., Kosovichev, A.~G., Mansour, N.~N.\ 2009, \apj, 702, 1150


\bibitem[Zhao et al.(2012)]{2012ApJ...749L...5Z} Zhao, J., Nagashima, K., Bogart, R.~S., Kosovichev, A.~G., Duvall Jr., T.~L.\ 2012, \apjl, 749, 5


\bibitem[Zhao et al.(2013)]{2013ApJ...774L..29Z} Zhao, J., Bogart, R.~S., Kosovichev, A.~G., Duvall Jr., T.~L.,  Hartlep, T.\ 2013, \apjl, 774, 29



\end{thebibliography}
\end{document}